\documentclass[]{IEEEtran}
\usepackage{cite}
\usepackage{amsmath,amssymb,amsfonts}
\usepackage{algorithmic}
\usepackage{graphicx}
\usepackage{textcomp}

\newtheorem{theorem}{Theorem}

\newtheorem{remark}{Remark}

\newcommand{\carrew} {\hfill $\Box$}

\newcommand{\carre}{\begin{flushright} \rule{2mm}{2mm} \end{flushright}}
\def\BibTeX{{\rm B\kern-.05em{\sc i\kern-.025em b}\kern-.08em
		T\kern-.1667em\lower.7ex\hbox{E}\kern-.125emX}}

\begin{document}

	\title{
Reduction of Velocity-Dependent Terms in Total Energy Shaping Approach}
\author{ M. Reza J. Harandi and Mehrzad Namvar
}

\maketitle
\begin{abstract}
Total energy shaping through interconnection and damping assignment passivity-based control (IDA-PBC) provides a powerful and systematic framework for stabilizing underactuated mechanical systems. 
Despite its theoretical appeal, incorporating actuator limitations into total energy shaping remains a largely open problem, with only limited results reported in the existing literature. 
In practice, the closed-loop behavior of energy-shaping controllers is strongly affected by the kinetic energy shaping terms.
In this paper, a simultaneous IDA-PBC (SIDA-PBC) framework is employed to systematically attenuate the kinetic energy shaping terms by exploiting generalized forces, without altering the matching partial differential equations (PDEs). 
The free component of the generalized forces is derived analytically via an $\ell_\infty$-norm optimization formulation. 
Although a reduction in kinetic energy shaping terms does not necessarily guarantee a decrease in the overall control effort, the proposed approach effectively suppresses kinetic energy shaping components and achieves a reduced control magnitude whenever such a reduction is structurally feasible. 
Unlike existing approaches based on gyroscopic terms, which require multiple actuators, the proposed method is applicable to mechanical systems with a single actuator. 
Simulation and experimental results are provided to validate the effectiveness of the proposed approach.

\end{abstract}
\section{INTRODUCTION}\label{s1}
Most existing methods for stabilizing underactuated systems, i.e., systems with fewer actuators than degrees of freedom~\cite{liu2019antiswing}, are largely case-specific~\cite{makki2024design}. In contrast, total energy shaping via interconnection and damping assignment passivity-based control (IDA-PBC) provides a general and systematic framework for stabilization by assigning a desired closed-loop Hamiltonian function~\cite{ortega2002stabilization}.
Unlike fully actuated systems, for which bounded-input stabilization has been extensively studied~\cite{harandi2023practical}, incorporating actuator limitations into total energy shaping remains scarcely addressed for robotic systems~\cite{lu2019nonlinear}. A key difficulty arises from the kinetic energy shaping terms that appear explicitly in the IDA-PBC control law, while such terms are absent in potential energy shaping.

In classical passivity-based control (PBC), the potential energy is shaped to stabilize the desired equilibrium by exploiting the inherent passive structure of the system~\cite{haghjoo2025unified}. IDA-PBC extends this framework by explicitly prescribing the closed-loop interconnection and damping structure and computing the assignable Hamiltonian functions through the solution of matching partial differential equations (PDEs)~\cite{ortega2002stabilization,afkar2024decentralized}. Due to the complexity of these PDEs, substantial research efforts have focused on their solvability and reformulation~\cite{romero2016energy,harandi2023reformulation}. More recently, robust and adaptive variants of IDA-PBC have also been proposed~\cite{franco2024integral,harandi2024adaptive,he2019design}; see~\cite[Ch.~2]{harandi2021passivity} for a comprehensive overview. Moreover, simultaneous energy shaping and damping injection, as a generalized form of IDA-PBC, has been introduced in~\cite{donaire2016simultaneous}. Nevertheless, the problem of total energy shaping under actuator constraints has received very limited attention.

A few attempts have been made to address input limitations within energy shaping controllers. In~\cite{santibanez2005control}, a bounded-input total energy shaping controller was developed for the inertia wheel pendulum by eliminating kinetic energy shaping terms, thereby avoiding unbounded control actions. In~\cite{j2021bounded}, the IDA-PBC control law was analyzed term by term, and worst-case upper bounds were derived for each component, allowing an overall bound to be obtained. Additionally, appropriate designs of the desired potential energy were proposed to confine homogeneous terms. Another line of research exploits the free components of the desired interconnection matrix.
In particular, \cite{harandi2023stabilization} used the free skew-symmetric (gyroscopic) terms of the interconnection matrix and reduced the magnitude of the control law. However, this approach is limited to systems with at least two actuators, whereas most benchmark underactuated systems possess only a single actuator.

In this paper, we focus on minimizing the kinetic energy shaping terms in the control law of energy shaping approach for robotic systems.To this end, we employ simultaneous IDA-PBC (SIDA-PBC), a more general framework in which generalized forces are incorporated instead of relying solely on gyroscopic terms.
In contrast to~\cite{harandi2023stabilization} that uses the free part of gyroscopic terms, the aim is to design these generalized forces to minimize the norm of kinetic energy shaping components in control law without modifying the PDEs.
For this purpose, the design procedure is carried out by analytically determining the free part of the generalized forces matrix using an \(\ell_\infty\)-norm optimization strategy.
Consequently, the suggested approach can also be applied to underactuated systems with a single actuator while avoiding undesirable fractional terms; see \textit{Remark~\ref{re2}} for more details. 
However, since the optimization of kinetic energy shaping terms does not necessarily decrease the magnitude of overall control effort, the optimization should be applied conditionally; see \textit{Remark~\ref{re3}} for a detailed explanation.
Simulation results on the Pendubot, together with experimental validation on a 3-DOF robot, verify the effectiveness of the proposed approach. 

The main contributions of this paper are summarized as follows:
\begin{itemize}
	\item A systematic method is proposed to reduce kinetic energy shaping terms in the control law of mechanical systems within the SIDA-PBC framework, without modifying the matching PDEs.
	\item An $\ell_\infty$-norm optimization problem is formulated with respect to the free part of the generalized forces matrix, and an analytical solution is derived to minimize the magnitude of kinetic energy shaping components in the control input.
	\item Unlike prior results based on gyroscopic terms, the proposed approach yields a smooth control law by avoiding fractional terms in the control input, thereby improving practical implementability, and is applicable to mechanical systems with a single actuator.
	\item Experimental validation on a haptic system indicates that attenuating the kinetic energy shaping terms yields an approximate $30\%$ reduction in the peak control effort.
\end{itemize}
These contributions collectively extend the applicability of total energy shaping control under actuator constraints while preserving the structural properties of IDA-PBC.

\section{Overview of SIDA-PBC}\label{sec:21}
A mechanical system may be represented by the following equations~\cite{donaire2016simultaneous}
\begin{equation}
	\label{1}
	\begin{bmatrix}
		\dot{q} \\ \dot{p}
	\end{bmatrix}
	=\begin{bmatrix}
		0_{n\times n} & I_n \\ -I_n & 0_{n\times n}
	\end{bmatrix}
	\begin{bmatrix}
		\nabla_q H \\ \nabla_p H
	\end{bmatrix}
	+\begin{bmatrix}
		0_{n\times m} \\ G
	\end{bmatrix}
	u
\end{equation}
where $q\in \mathbb{R}^n$ represent the position, $p\in \mathbb{R}^n$ is momentum in which $p=M(q)\dot{q}$, $H(q,p)=\frac{1}{2}p^\top M^{-1}(q)p+V(q)$ represents the Hamiltonian, consisting of kinetic and potential energy.  $0<M(q)$ is
the inertia matrix and $G(q)\in \mathbb{R}^{n\times m}$  is input mapping matrix. By applying
\begin{align}
	&u=G^\dagger\big(\nabla_q H-M_dM^{-1}\nabla_q H_d+\Lambda M_d^{-1}p \big)
	\label{4}
\end{align}
to system (\ref{1})
with 
$$H_d=\frac{1}{2}p^\top M_d^{-1}(q)p+V_d(q),$$ and $$G^\dagger=(G^\top G)^{-1}G^\top,$$
in which $M_d$ and $V_d$ are required to fulfill the following partial differential equations
\begin{align}
	&G^\bot \{\nabla_q  \big(p^\top M^{-1}(q)p\big) - M_dM^{-1}\nabla_q \big(p^\top M_d^{-1}(q)p\big)\nonumber\\   & +2\Lambda M_d^{-1}p \} =0,   \label{5} \\&
	G^\bot \{\nabla_q V-M_dM^{-1}\nabla_q V_d\} =0, \label{6}
\end{align}
the equations of closed-loop system will be
\begin{equation}
	\label{2}
	\begin{bmatrix}
		\dot{q} \\ \dot{p}
	\end{bmatrix}
	=
	\begin{bmatrix}
		0_{n\times n} & M^{-1}M_d \\ -M_dM^{-1} & \Lambda
	\end{bmatrix}
	\begin{bmatrix}
		\nabla_q H_d \\ \nabla_p H_d
	\end{bmatrix},
\end{equation}
in which $\Lambda M_d^{-1}p$ is generalized force and $\Lambda(q,p)\in\mathbb{R}^{n\times n}$ should satisfy $\Lambda+\Lambda^\top\preceq 0$. Note that a particular case is $\Lambda=J_2(q,p)-GK_vG^\top$ where $J_2=-J_2^\top$ and $0<K_v$ is damping term. This structure corresponds to the standard IDA-PBC formulation~\cite{ortega2002stabilization}. 
Notice that the desired equilibrium point denoted by $q^*$ which satisfy $G^\bot \nabla_q V=0$ should be minimum of 
$V_d$. 
Then $q^*$ is stable since the derivative of $H_d$ as a Lyapunov function is
\begin{align}\label{sta}
	(\nabla_p H_d)^\top(\Lambda +\Lambda^\top)\nabla_p H_d\leq0.
\end{align}
Note that, similar to other studies on the development of IDA-PBC~\cite{harandi2024adaptive,harandi2023stabilization,franco2024integral}, it is assumed that the solutions of~(\ref{5}) and~(\ref{6}) have been derived beforehand. 
Moreover, when the system is fully actuated, the PDEs~(\ref{5}) and~(\ref{6}) are omitted, and the desired inertia and potential energy functions can be selected freely~\cite{harandi2023stabilization}.

\section{Main Results}
The control law (\ref{4}) consists of kinetic energy shaping, potential energy shaping, and damping terms. Because it is not possible to inject damping into the unactuated coordinates, the damping term in the control law necessarily takes the form $K_vG^{\top}\nabla_p H_d$~\cite{harandi2024adaptive}. Hence,
\begin{align}
	\label{lmb}
	\Lambda=\Lambda_{k}+\Lambda_{u}-GK_vG^\top,
\end{align}
where the negative semi-definite matrix $\Lambda_{k}$ is employed to solve the kinetic energy PDE, and $\Lambda_{u}$ is a design parameter that will be determined later.
Following~\cite{j2021bounded}, the damping injection term can also be redesigned as $K_v\tanh(G^{\top}\nabla_p H_d)$ with a sufficiently small gain. Moreover, \cite{j2021bounded} proposes a design method for the homogeneous part of $V_d$ such that the potential energy term remains bounded. Consequently, the main difficulty arises from the kinetic energy terms, expressed as
\begin{align}
	u_{ki}=G^\dagger\big(\nabla_q K-M_dM^{-1}\nabla_q K_d+\Lambda_{k}\nabla_p H_d\big),\label{uki}
\end{align}
where $K$ and $K_d$ denote the actual and desired kinetic energy, respectively. As discussed in section~\ref{s1}, the objective is now to design $\Lambda_{u}$ in a way that minimizes the norm of $u_{ki}$.
However, the matrix $\Lambda_{u}$ may appear in PDE (\ref{5}) and consequently alter its previously derived solution. To avoid resolving the nonlinear and analytically intractable PDE, it is reasonable to constrain the structure of $\Lambda_{u}$ as
$$\Lambda_{u}=G\Lambda_{ua},$$ where $ \Lambda_{ua}\in\mathbb{R}^{m\times n}$.
This directly implies 
$$G^\bot\Lambda_{u}=G^\bot G\Lambda_{ua}=0,$$ thereby preserving the solution of the kinetic energy matching PDE. Moreover, according to the stability condition (\ref{sta}), the matrix $\Lambda$ must be negative semi-definite. To satisfy this requirement, $\Lambda_{u}$ is defined as
$$\Lambda_{ua}=\Lambda_{uan}G^\top,\quad \to\quad \Lambda_{u}=G\Lambda_{uan}G^{\top},$$
where the design matrix $\Lambda_{uan} \in \mathbb{R}^{m \times m}$ should satisfy
\begin{align}\label{con}
	\Lambda_{uan}+\Lambda_{uan}^{\top}\preceq 0.
\end{align}
This selection is justified by noting that, based on (\ref{lmb}), a sufficient condition for ensuring $\Lambda+\Lambda^{\top}\preceq 0$ is $	\Lambda_{u}+\Lambda_{u}^\top\preceq 0$ which directly yields
$$	G({\Lambda_{uan}+\Lambda_{uan}^{\top}})G^\top\preceq 0.$$

%
%
%

Now, we may represent the main problem as follows
\begin{align}
	\min |u_{ki_i}+(\Lambda_{uan}G^\top\nabla_p H_d)_i|,\quad \; i\in\{1,\dots,m\} \label{cn}
\end{align}
such that (\ref{con}) is satisfied in which $*_i$ denotes $i$-th element of $*$. Similar to \cite{harandi2023stabilization}, (\ref{cn}) may be represented as a minimization of \(\ell_\infty\)-norm.
Therefore, the objective is to solve the following optimization problem
\begin{align}
	\min_{\mbox{s. to} \;\Lambda_{uan}+\Lambda_{uan}^\top\preceq 0} \|u_{ki}+\Lambda_{uan}G^\top M_d^{-1}p\|_\infty\label{min}.
\end{align}
To solve (\ref{min}), the following theorem is required.

\begin{theorem}\label{thm:linf_opt}\normalfont
	Consider the following optimization problem
	\begin{align}\label{opt}
		 &\min_{A}\; \|A x - b\|_\infty,\\
		 &\mbox{s. to}\; A \preceq 0 \nonumber
	\end{align}
	in which $A\in\mathbb{R}^{n\times n}, x\in\mathbb{R}^n, b\in\mathbb{R}^n$ and $x\neq 0$.
The solution of (\ref{opt}) is $A^*=A_{s}^*+A_w^*$ in which the symmetric matrix $A_{s}^*$ and the skew-symmetric matrix $A_w^*$ are as follows
\begin{subequations}\label{eq:AwAs}
	\begin{align}
		A_{s}^* &= \frac{a_s^*}{\|x\|_2^2}\, I_n, 
		& a_s^* &= \min\{0, x^\top b\},
		\label{s}
		\\[4pt]
		A_{w}^* &= \frac{v^* x^\top - x {v^*}^\top}{\|x\|_2^2}, 
		& v^* &= b - A_{s}^* x - \xi^*, 
		\label{w}
	\end{align}
\end{subequations}
where the $i$-th element of $\xi^*\in\mathbb{R}^n$ is:
\begin{align*}
\xi_i^* = \operatorname{sign}(x_i)\;
	\frac{x^\top b - a_s^*}{\|x\|_1}.
\end{align*}
\end{theorem}

\hfill $\square$

Recall that for a square matrix $A$, there always exists unique matrices $A_s$ and $A_w$ such that $A=A_s+A_w$ where $A_s$ is symmetric and $A_w$ is skew-symmetric.

\textbf{proof}.
To prove \textit{Theorem~\ref{thm:linf_opt}}, first, we compute $A_w^*$ based on $A_s$ and then,  $A_s^*$ will be designed. Thus,
consider the fixed symmetric matrix \(A_s \preceq 0\) and define the residual vector $r$ as follows
$$	r:= b - A_sx.$$
Since \(A_w=-A_w^\top\), the vector 
$$v := A_wx,$$
 satisfies
$	x^\top v = 0$.
Hence, the internal optimization based on \(A_w\) will be in the following form
\begin{align}\label{o}
	\min_{x^\top v = 0} \|v - r\|_\infty.
\end{align}
By defining
$$\xi := r - v,$$
the optimization problem (\ref{o}) is equivalent with
\begin{align}\label{ot}
		\min_{\xi^\top x = r^\top x} \|\xi\|_\infty,
\end{align}
since
$$\xi^\top x = r^\top x-v^\top x=r^\top x.$$
Since the \(\ell_\infty\)-norm bounds the absolute value of each component, it is clear that the absolute values of the elements of the above optimization denoted by $\xi^*$ are equal~\cite{harandi2023stabilization}. Furthermore, by the triangle inequality we have
\begin{align*}
	|r^\top x|& = |\xi^\top x| = \left|\sum_i \xi_i x_i \right| 
\leq \sum_i |\xi_i| |x_i|	\\& \leq \max_i |\xi_i|\sum_i  |x_i|= \|\xi\|_\infty \|x\|_1.
\end{align*}
Thus, the upper bound of $	\|\xi\|_\infty$ is
\begin{align*}
	\|\xi\|_\infty\geq	\frac{|r^\top x|}{\|x\|_1}.
\end{align*}
If $\xi$ is designed such that the absolute values of its elements are equal and the above bound is attainable, it will be the solution of (\ref{ot}).
Hence,
\begin{align*}
	\xi_i^* =  \operatorname{sign}(x_i) \frac{|r^\top x|}{\|x\|_1} ,
\end{align*}
which apparently satisfies \(\xi^\top x = r^\top x\),  \(\|\xi\|_\infty = \frac{|r^\top x|}{\|x\|_1}\), and 
$$|\xi_i|=\xi_j|\qquad \mbox{for every}\; i,j\in\{1,\dots,n\}.$$
So far, $\xi^*$ and consequently, $v^*$ and $A_w^*$ have been determined based on $A_s$ which lies inside $r$. The next step is to derive $A_s^*$.

To compute $A_s^*$, we need to find the minimizer of $|r^\top x|$.
Since
$	r^\top x = b^\top x - x^\top A_s x,$
and also 
$$A_s \preceq 0,\qquad\to \qquad	a_s := x^\top A_s x \leq 0,$$
the aforementioned minimization problem becomes
\begin{align*}
	\min_{a_s \leq 0} {|b^\top x - a_s|}.
\end{align*}
The trivial solution of the above minimization is
\begin{align*}
	a_s^* = \min\{b^\top x, 0\},
\end{align*}
yielding the following optimal objective value for the problem (\ref{opt}):
	\begin{align}
		\Phi^*:=
			\begin{cases}
					0, & b^\top x \le 0, \\[6pt]
				\frac{(b^\top x)}{\|x\|_1}, & b^\top x > 0.
				\end{cases}\label{phi}
		\end{align}
Finally, the optimal matrices may be constructed simply as (\ref{eq:AwAs}).	
This concludes the proof.
\carre
\begin{remark}\label{re1}\normalfont
	The optimizer $A^*=A_s^*+A_w^*$ is not unique. Indeed, for any fixed $a_s^*$ and $v^*$ satisfying $v^{*\top}x = 0$, there exist infinite symmetric matrices $A_s$ such that $x^\top A_s x = a_s^*$, and likewise infinite skew-symmetric matrices $A_w$ satisfying $A_w x = v^*$.
	To obtain a unique representative, additional selection criteria may be imposed. In this work, we choose $A_s^*$ and $A_w^*$ to be smooth functions of $x$. \carrew
\end{remark}

Now, to solve the optimization problem (\ref{min}), it is sufficient to apply \textit{Theorem~\ref{thm:linf_opt}} with the following parameters
$$A=\Lambda_{uan},\qquad x=G^\top M_d^{-1}p,\qquad b=-u_{ki}.$$
Notice that the case $x=G^\top M_d^{-1}p=0$ implies either $p=0$ or $M_d^{-1}p\in\mathrm{null}(G^\top)$. In the former case, there is no need to optimize kinetic energy shaping terms since $u_{ki}=0$.
In the latter case, it is not possible to modify (\ref{uki}) via $\Lambda_{uan}$. For the sake of clarity, the overall kinetic energy shaping term is defined as
\[
u_{ov-ki} = Ax - b = \Lambda_{uan} G^\top M_d^{-1} p + u_{ki},
\]
which combines the nominal and additional kinetic energy shaping terms.

Note that if $b^\top x=-u_{ki}^\top G^\top M_d^{-1}\leq 0$, the control law does not include kinetic energy shaping terms. Although the interpretation of this condition is not straightforward, it is noteworthy that during the transient response, the components related to kinetic energy shaping within the control law are sometimes entirely omitted. Furthermore, if $b^\top x=-u_{ki}^\top G^\top M_d^{-1}> 0$, it follows that $A_s=0$ and merely the skew-symmetric matrix $A_w$ is designed. In this case, $\Phi^*$ given in (\ref{phi}) coincides with the value reported in equation (23) of \cite{harandi2023stabilization}. However, here, the optimal $A_w^*$ is directly obtained from (\ref{w}), whereas in \cite{harandi2023stabilization}, its derivation requires the inverse of a sub-block of a matrix that leads to fractional terms; see the following remark.

\begin{remark}\label{re2}\normalfont
	In \cite{harandi2023stabilization}, gyroscopic terms were employed to reduce the norm of the control law. However, this method is only applicable to mechanical systems with at least two actuators and cannot be applied to systems with a single actuator, due to the nonexistence of a nonzero $1\times 1$ skew-symmetric matrix. Consequently, its applicability is limited to a small subset of underactuated systems, as most benchmark examples involve only a single actuator.
	 Furthermore, to design this matrix, its elements are first collected in the vector $\theta$ and the problem is formulated as $\|Y\theta-b\|_\infty$ in which $Y$ is proportional to $p$.
	The solution is then given by $\theta^\star=Y^\top x^{o}$ in which $x^{o}$ is proportional to $\bar{A}_e^{-1}$ with $\bar{A}=YY^\top$ and $\bar{A}_e$ is obtained by removing $\ell$-th row and column of $A$. This determination may lead to large values in the skew-symmetric matrix due to fractional terms, thereby degrading the controller's performance. In this work, we employ generalized forces and propose a novel solution to the optimization problem that addresses the aforementioned issues. \carrew
\end{remark}
\begin{remark}\label{re3}\normalfont
	In this paper, we focus on the kinetic energy shaping terms.
	Although we optimize the norm of $u_{ov-ki}$, this does not necessarily lead to reduction in $\|u\|$, specially when the sign of $u_{ov-ki}$  differs from the sign of other components of $u$. Accordingly, the final control law applied to the system is designed as follows
	\begin{align}\label{red}
		u=\min\{u_{IDA},u_{Th1-IDA}\}
	\end{align}
	in which $u_{IDA}$ refers to (\ref{4}), and $u_{Th1-IDA}$ denotes the control law of IDA-PBC  together with Theorem~\ref{thm:linf_opt}. \carrew
\end{remark}
In the next section, we apply the proposed optimization method to a benchmark system to evaluate its performance. It is worth noting that, using the proposed method, the magnitude of 
$u$ is reduced without imposing a significant computational burden and without violating the stability conditions. This demonstrates the effectiveness of the proposed approach, which can be readily applied to any robotic system. 

\begin{figure}[]
	\centering
	\includegraphics[scale=.3]{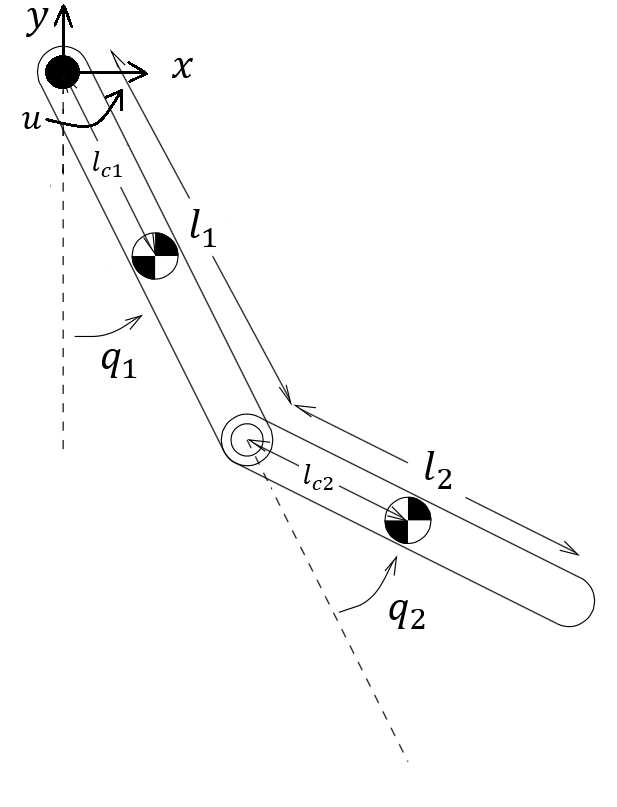}
	\caption
	{Schematic of the Pendubot.}
	\label{sh}
\end{figure}
\section{Simulation Results}
To assess the effectiveness of the proposed method, we apply it to Pendubot system through simulation. The robot illustrated in Fig.~\ref{sh} is a two-DOF system in which only the first joint is actuated.
 The dynamical terms of this robot are given as follows~\cite{sandoval2008interconnection,harandi2022solution}
\begin{align*}
	&	M=\begin{bmatrix}
		c_1+c_2+2c_3\cos(q_2) & c_2+c_3\cos(q_2) \\  c_2+c_3\cos(q_2) & c_2
	\end{bmatrix},\\& G=\begin{bmatrix}
		1 \\ 0
	\end{bmatrix},
	V=-c_4g\cos(q_1)-c_5g\cos(q_1+q_2),
\end{align*}
in which
\begin{align*}
	c_1 &= I_1+ l_{c_1}^2m_1 +  l_1^2m_2 , \qquad
	c_2 =  I_2+ l_{c_2}^2m_2, \\
	c_3 &=  l_1 l_{c_2}m_2, \qquad 
	c_4 =  l_{c_1}m_1 +  l_1m_2, \qquad
	c_5 =  l_{c_2}m_2.
\end{align*}
In~\cite{sandoval2008interconnection}, an IDA-PBC was designed for this system.
Here, the robot parameters and controller components are selected to be consistent with those in \cite{sandoval2008interconnection}. For clarity, these quantities are restated as follows
\begin{align*}
	&M_d = k_3
	\begin{bmatrix}
		\rho & c_1 - c_2 \\
		c_1 - c_2 & -c_2 + c_3 \cos(q_2)
	\end{bmatrix},\\&
	V_d = \frac{c_5 g}{k_3}\big[\cos(q_1 + q_2) + 1\big]
	+ \frac{k_p}{2}\big[q_2 + 2q_1 - 2\pi \big]^2.
\end{align*}
\begin{figure}[t]
	\centering
	\includegraphics[trim={0.75cm 0.2cm .7cm 0.3cm},clip,width=8.cm,height=7.5cm]{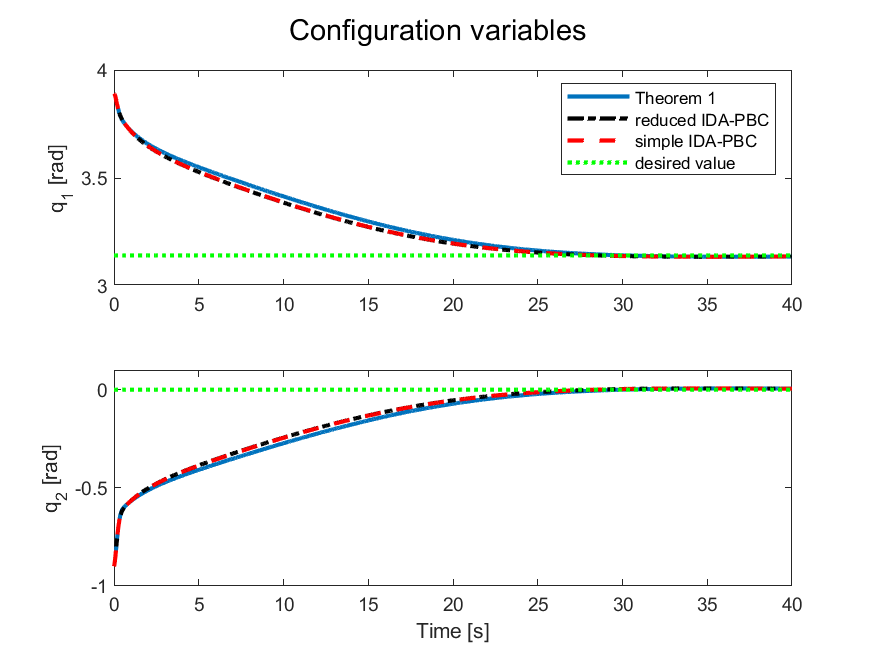}
	\caption
	{Configuration variables of the Pendubot under the three controllers. Both $q_1$ and $q_2$ converge toward the reference values corresponding to the upright position.}
	\label{sh1}
\end{figure}
\begin{figure}[]
	\centering
	\includegraphics[scale=.6]{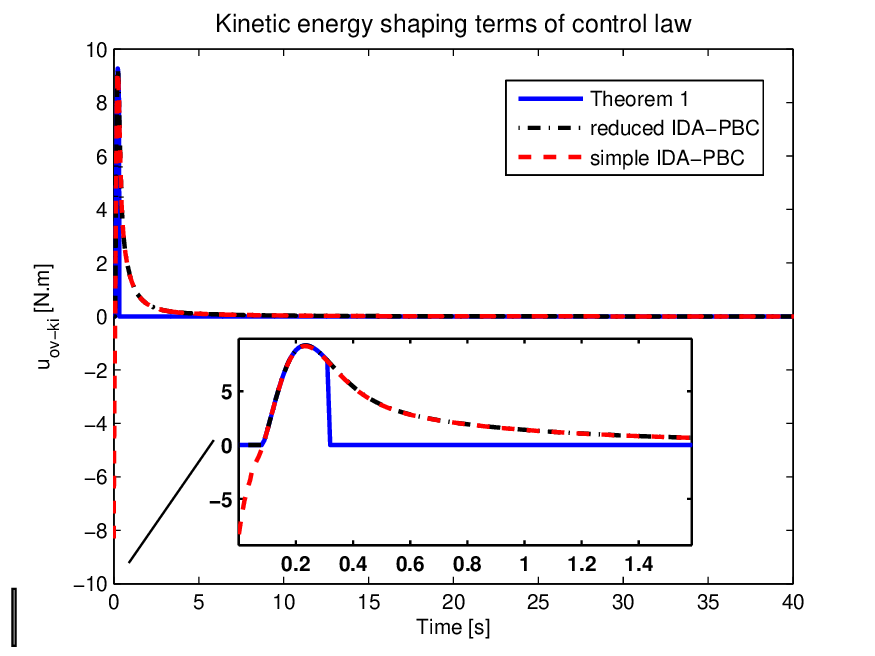}
	\caption
	{Comparison of the kinetic energy shaping terms for the three controllers. According to Theorem~\ref{thm:linf_opt}, $u_{ov-ki}$ remains almost zero throughout the response, whereas under (\ref{red}) it is zero only during the initial moments. }
	\label{sh2}
\end{figure} 
\begin{figure}[]
	\centering
	\includegraphics[scale=.6]{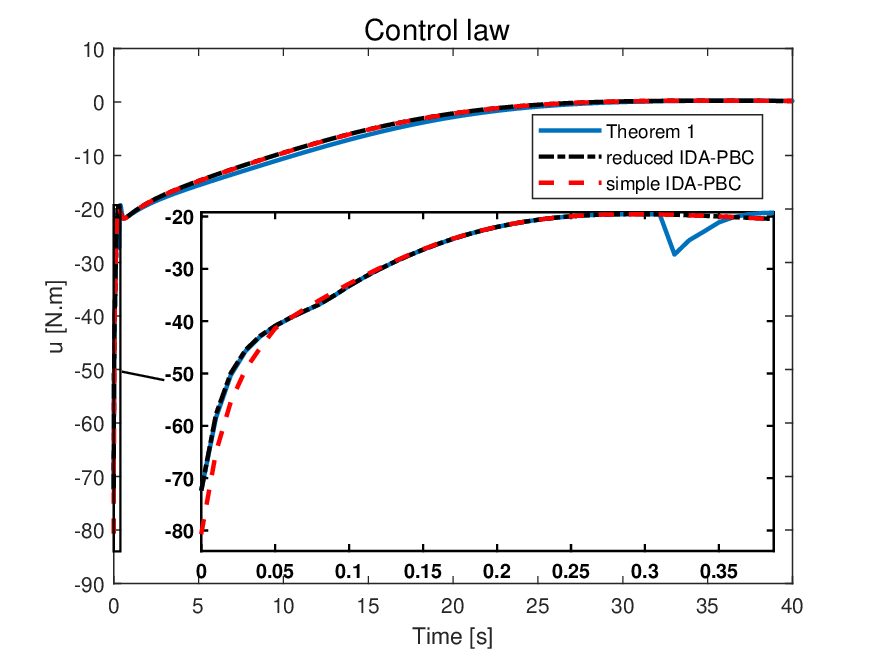}
	\caption
	{Control signals generated by the three controllers. The control input resulting from the control law in (\ref{red}) exhibits a smaller magnitude compared to the others.}
	\label{sh3}
\end{figure} 
%
%

Three controllers are applied to the robot: the simple IDA-PBC given in (\ref{4}), the IDA-PBC with minimized kinetic energy shaping terms based on \textit{Theorem~\ref{thm:linf_opt}}, and the controller described in \textit{Remark~\ref{re3}} and presented in (\ref{red}), which will be referred to as the "reduced IDA-PBC" in the sequel. The results are shown in Figs.~\ref{sh1}-\ref{sh3}. The configuration variables of the robot are depicted in Fig.~\ref{sh1}, where the robot converged to the upright position and the transient responses were similar across different control laws.
However, Fig.~\ref{sh2} illustrates that the kinetic energy shaping terms differ across the controllers. Using \textit{Theorem~\ref{thm:linf_opt}}, $u_{ov-ki}$ becomes nonzero only within the time interval approximately $(0.1,0.35)$, whereas for the reduced IDA-PBC, $u_{ov-ki}$ remains zero in the interval approximately $\approx (0,0.1)$. This observation may suggest that the controller designed using \textit{Theorem~\ref{thm:linf_opt}} provides the best performance. Nevertheless, to make a fair comparison, it is necessary to consider the control efforts shown in Fig.~\ref{sh3}. Evidently, the minimum control effort among the three controllers corresponds to the reduced IDA-PBC given in (\ref{red}). At the initial time, the control signals of \textit{Theorem~\ref{thm:linf_opt}} and (\ref{red}) are around $-70\;\text{N·m}$, while the simple IDA-PBC reaches approximately $-80\;\text{N·m}$. This demonstrates the superiority of the proposed method, as the peak of $u$ is reduced by approximately $13\%$ via an analytic optimization that imposes no significant computational burden. However, the absolute value of the control law of \textit{Theorem~\ref{thm:linf_opt}} increases at $t \approx 0.35$ since, according to Fig.~\ref{sh2}, $u_{ov-ki}$ is positive at this time, and eliminating the kinetic energy terms from $u$ leads to an increase in $|u|$. This observation supports the discussion in \textit{Remark~\ref{re3}}, which states that \textit{Theorem~\ref{thm:linf_opt}} should only be applied when it leads to a reduction in the absolute value of the control signal.
\begin{figure}[t]
	\centering
	\includegraphics[scale=.54]{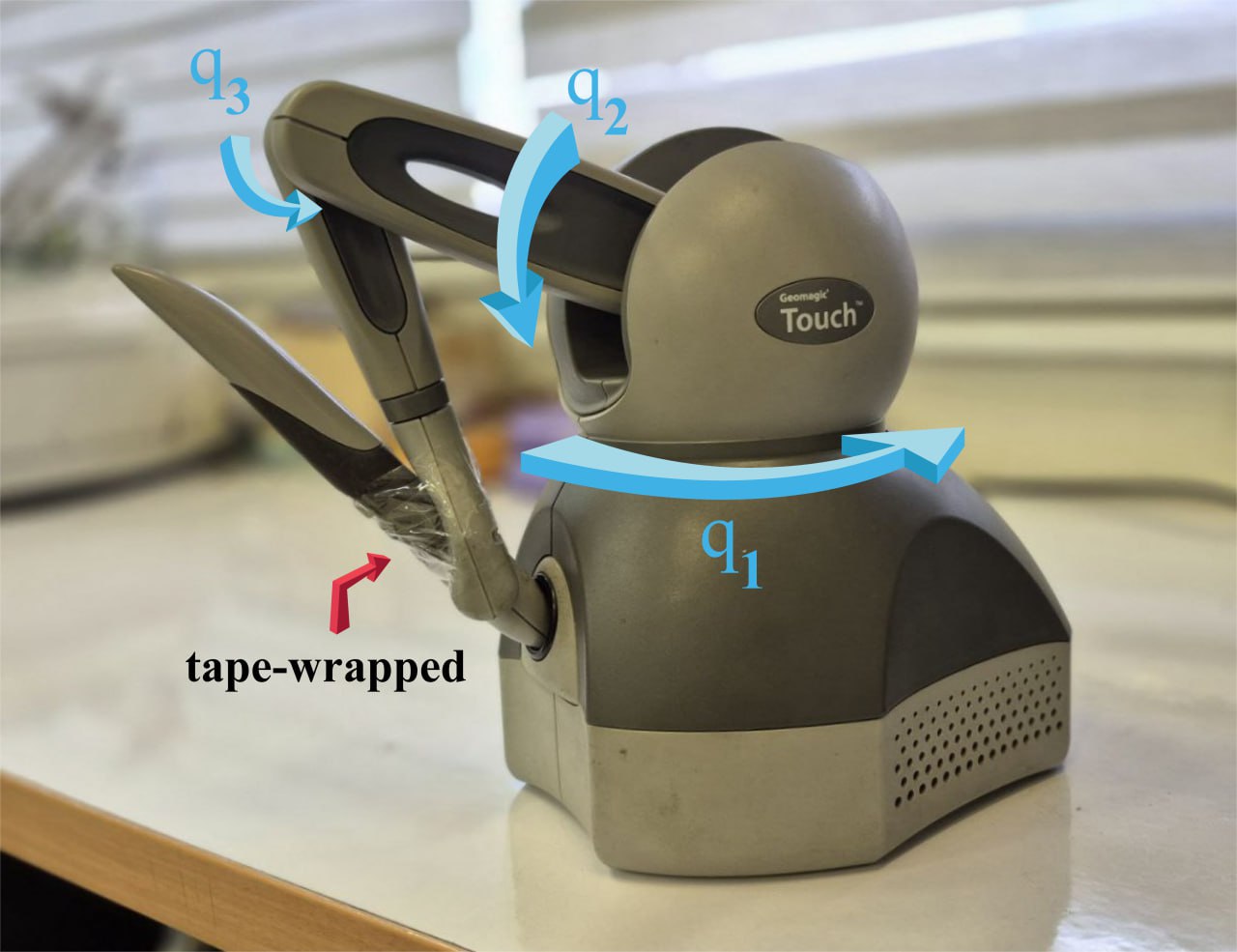}
	\caption
	{Schematic of the Geomagic Touch robot.}
	\label{p}
\end{figure}
\section{Experimental Results}
\begin{figure}[]
	\centering
	\includegraphics[scale=.6]{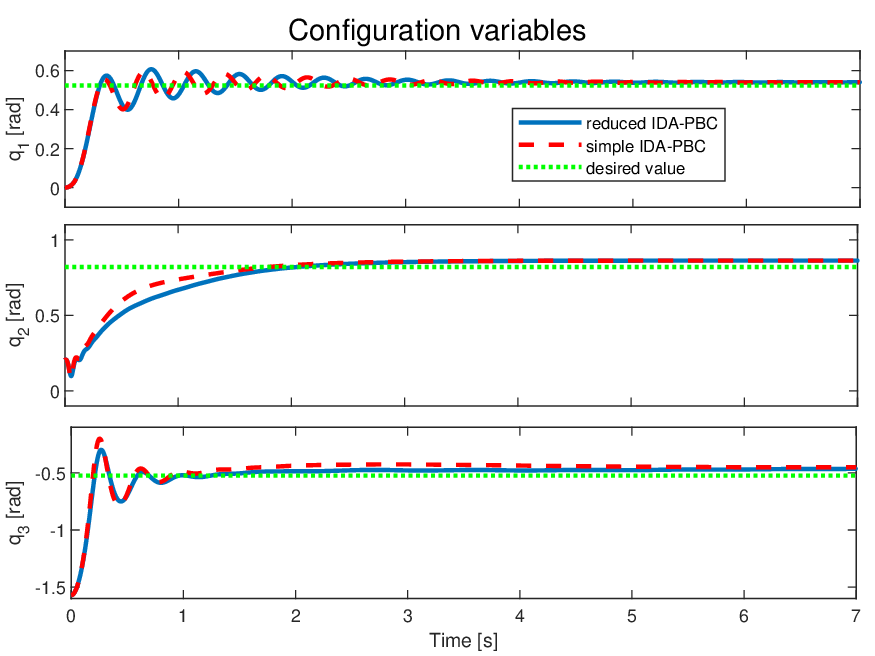}
	\caption
	{Configuration variables of the robot, illustrating that the controllers exhibit similar steady-state errors.}
	\label{p1}
\end{figure}

\begin{figure}[h]
	\centering
	\includegraphics[scale=.6]{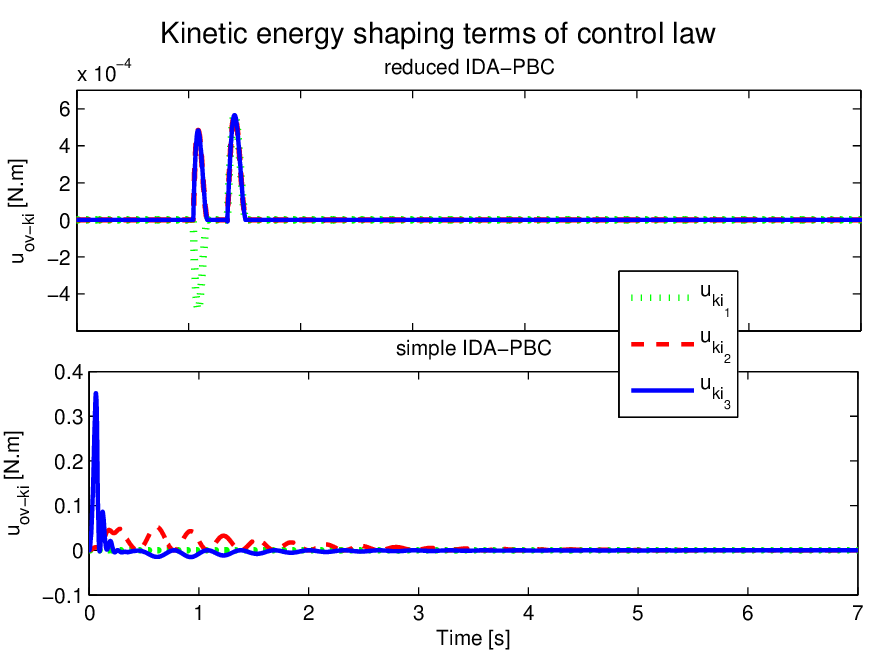}
	\caption
	{Kinetic energy shaping terms of the controllers, showing that the application of Theorem~\ref{thm:linf_opt} results in a negligible contribution of $u_{ov-ki}$. }
	\label{p2}
\end{figure}

\begin{figure}[]
	\centering
	\includegraphics[scale=.6]{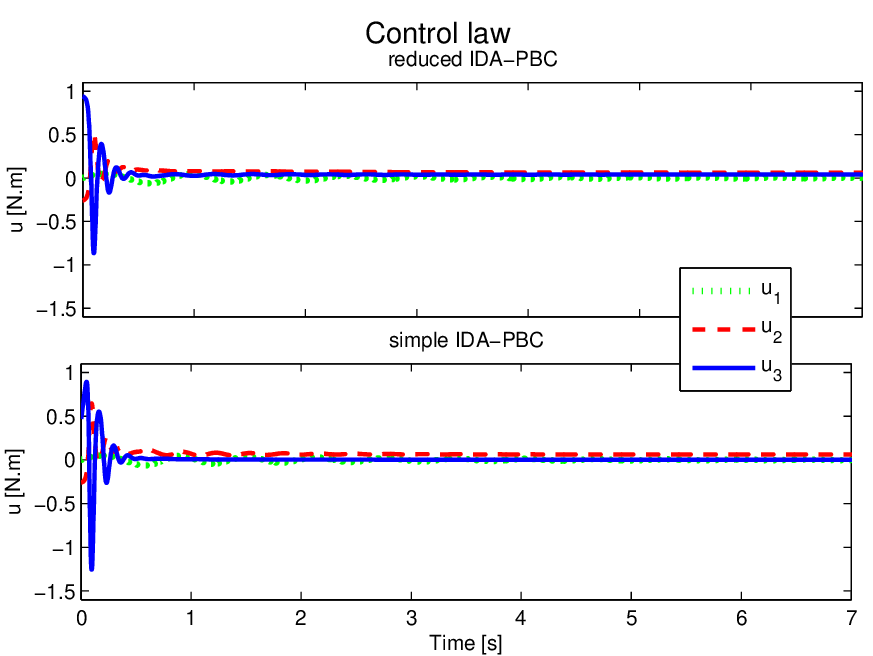}
	\caption
	{Control signals of the controllers, demonstrating that the proposed method reduces the peak control effort.}
	\label{p3}
\end{figure}

The experimental platform used to validate the proposed control approach is shown in Fig.~\ref{p}. 
It consists of a fully actuated 3-DOF Geomagic Touch\texttrademark\ haptic interface. 
The controller is implemented in MATLAB/Simulink and integrated through a C-MEX S-function, which enables communication with a C++ application developed using OpenHaptics\textregistered. 
The closed-loop control is executed at a sampling frequency of $1\,\mathrm{kHz}$. 
The robot dynamics are expressed as follows:
\begin{align*}
&M_{11}=\phi_1 \cos^2(q_2) + \phi_2 \cos(q_2)\cos(q_2+q_3)+ \\& \phi_3 \sin^2(q_2+q_3), \quad M_{12}=M_{13}=0,\quad  M_{22}=\phi_1+ \\& 2\phi_2 \cos(q_3) + \phi_3,\quad M_{23}=\phi_2 \cos(q_3) + \phi_3, \quad M_{33}=\phi_3,\\&
	V(q) = g \left( \phi_4 \sin(q_2) + \phi_5 \sin(q_2+q_3) \right),
\end{align*}
and the nominal values of the corresponding dynamic parameters are provided in Table~\ref{tab:dynamic_parameters}~\cite{kameli2025variable}.
\begin{table}[b]
	\centering
	\caption{Nominal values of the dynamic parameters.}
	\label{tab:dynamic_parameters}
	\begin{tabular}{cc}
		\hline
		\textbf{Parameter} & \textbf{Value} \\
		\hline
		$\phi_1$ & $0.00251729$ \\
		$\phi_2$ & $0.00108246$ \\
		$\phi_3$ & $0.00137408$ \\
		$\phi_4$ & $0.00449158$ \\
		$\phi_5$ & $0.00534505$ \\
		\hline
	\end{tabular}
\end{table}
An IDA-PBC with the following parameters is applied to the mechanism
\begin{align*}
	M_d=\kappa I_{3}, \quad V_d=\sum_{i=1}^{3}k_{p_i}\int_{0}^{q_i-q^*_i} \tanh(s)ds,\quad J_2=0.
\end{align*}
Note that, since the robot is fully actuated, the matching PDEs are omitted. 
The controller gains are selected as
\begin{align*}
	\kappa=0.001,\qquad k_{p_1}=k_{p_2}=k_{p_3}=1,\qquad K_d=0.3I_3.
\end{align*}
Both the simple IDA-PBC and the reduced IDA-PBC proposed in Remark~\ref{re3} are applied to the robot, and the corresponding results are illustrated in Figs.~\ref{p1}–\ref{p3}. 
The initial condition is chosen as $q_0^\top=[0,\pi/15,-\pi/2]$ with zero initial velocity, while the desired configuration is set to $q^*=[0.5,\pi/4,-0.5]^\top$.

As shown in Fig.~\ref{p1}, the configuration variables converge close to their desired values, and the transient responses of the controllers are relatively similar. 
However, the kinetic energy shaping terms, and consequently the control signals, exhibit significant differences. 
In particular, Fig.~\ref{p2} demonstrates that the term $u_{ov-ki}$ is negligible when the proposed method is employed, whereas its peak value reaches approximately $0.35~\text{N·m}$ under the simple IDA-PBC scheme. 
In this case, \textit{Theorem~\ref{thm:linf_opt}} is always applicable, since the matrix $A=\Lambda_{uan}\in\mathbb{R}^{3\times 3}$ in~(\ref{opt}) possesses a skew-symmetric part. 
Moreover, according to~(\ref{phi}), whenever $u_{ki}^\top G^\top M_d^{-1}p>0$, it follows that $u_{ov\text{-}ki}=0$.
Finally, as depicted in Fig.~\ref{p3}, the peak control effort is reduced by approximately $30\%$ due to the attenuation of the kinetic energy shaping terms. 
These results indicate that the proposed approach achieves satisfactory practical performance by reducing the control effort norm without imposing additional computational burden.

\section{Conclusion}
This paper presented a design-oriented framework for attenuating the kinetic energy shaping terms in passivity-based control laws. 
By exploiting generalized forces, a free negative semi-definite matrix was systematically constructed through the analytical solution of an $\ell_\infty$-norm optimization problem. 
A key insight of this study is that a reduction in the kinetic energy shaping terms does not, in general, guarantee a decrease in the overall control effort, which necessitates performing the optimization under a specific structural condition. 
This observation provides a deeper understanding of the relationship between energy shaping and control effort in IDA-PBC-based designs. 
The effectiveness of the proposed approach was demonstrated through simulations on the Pendubot and experimental validation on a haptic device, where a noticeable reduction in the peak control effort was achieved without increasing computational complexity.

\bibliographystyle{ieeetr}
\bibliography{reff}

@article{he2019design,
	title={Design and implementation of adaptive energy shaping control for DC--DC converters with constant power loads},
	author={He, Wei and Ortega, Romeo},
	journal={IEEE Transactions on Industrial Informatics},
	volume={16},
	number={8},
	pages={5053--5064},
	year={2019},
	publisher={IEEE}
}

@article{makki2024design,
	title={Design, Mathematical Modeling, and Control of an Underactuated 3-DOF Experimental Helicopter},
	author={Makki, Osamah Talal and Moosapour, Seyyed Sajjad and Mobayen, Saleh and Nobari, Jafar Heyrani},
	journal={IEEE Access},
	volume={12},
	pages={55568--55586},
	year={2024},
	publisher={IEEE}
}

@article{lu2019nonlinear,
	title={Nonlinear coordination control of offshore boom cranes with bounded control inputs},
	author={Lu, Biao and Fang, Yongchun and Sun, Ning},
	journal={International Journal of Robust and Nonlinear Control},
	volume={29},
	number={4},
	pages={1165--1181},
	year={2019},
	publisher={Wiley Online Library}
}

@article{kameli2025variable,
	title={A variable impedance control architecture for transparency improvement in nonlinear bilateral telerobotic systems},
	author={Kameli, Mobina and Motaharifar, Mohammad and Sayyaf, Negin and Heidari, Reza and Taghirad, Hamid D},
	journal={International Journal of Systems Science},
	pages={1--16},
	year={2025},
	publisher={Taylor \& Francis}
}

@article{haghjoo2025unified,
	title={Unified Passivity-Based Control for Turning and Walking in Asymptotically Stable Biped Robots},
	author={Haghjoo, Mohammad Reza},
	journal={IEEE Access},
	year={2025},
	publisher={IEEE}
}

@article{afkar2024decentralized,
	title={Decentralized control of DC microgrids using interconnection and damping assignment passivity-based control technique: Experimental verification},
	author={Afkar, Mohammad and Yuan, Cong and Gavagsaz-Ghoachani, Roghayeh and Saksiri, Wiset and Phattanasak, Matheepot and Martin, Jean-Philippe and Pierfederici, Serge},
	journal={IEEE Access},
	year={2024},
	publisher={IEEE}
}

@article{liu2019antiswing,
	title={An antiswing trajectory planning method with state constraints for 4-DOF tower cranes: design and experiments},
	author={Liu, Zhuoqing and Yang, Tong and Sun, Ning and Fang, Yongchun},
	journal={IEEE Access},
	volume={7},
	pages={62142--62151},
	year={2019},
	publisher={IEEE}
}

@article{franco2024integral,
	title={Integral ida-pbc for underactuated mechanical systems subject to matched and unmatched disturbances},
	author={Franco, Enrico and Arpenti, Pierluigi and Donaire, Alejandro and Ruggiero, Fabio},
	journal={IEEE Control Systems Letters},
	volume={8},
	pages={568--573},
	year={2024},
	publisher={IEEE}
}

@article{harandi2024adaptive,
	title={Adaptive total energy shaping of a class of manipulators},
	author={Harandi, M Reza J and Taghirad, Hamid D},
	journal={IET Control Theory \& Applications},
	volume={18},
	number={8},
	pages={1007--1015},
	year={2024},
	publisher={Wiley Online Library}
}

@article{harandi2023practical,
	title={Practical adaptive position feedback regulator for parallel robots with bounded inputs},
	author={Harandi, M Reza J and Hassani, A and Khalilpour, SA and Taghirad, Hamid D},
	journal={Journal of Vibration and Control},
	volume={29},
	number={11-12},
	pages={2634--2646},
	year={2023},
	publisher={SAGE Publications Sage UK: London, England}
}

@article{harandi2023reformulation,
	title={Reformulation of matching equation in potential energy shaping},
	author={Harandi, Mohammad Reza Jafari and Taghirad, Hamid D},
	journal={IEEE Transactions on Automatic Control},
	volume={68},
	number={12},
	pages={8275--8278},
	year={2023},
	publisher={IEEE}
}

@phdthesis{harandi2021passivity,
	title={Passivity Based Control of 3-DOF
Underactuated Suspended Cable-Driven
Robot},
author={Harandi, M Reza J},
year={2021},
school={ K. N. Toosi University of Technology}
}

@article{harandi2022solution,
	title={Solution of matching equations of IDA-PBC by Pfaffian differential equations},
	author={Harandi, M Reza J and Taghirad, Hamid D},
	journal={International Journal of Control},
	volume={95},
	number={12},
	pages={3368--3378},
	year={2022},
	publisher={Taylor \& Francis}
}

@article{romero2016energy,
	title={Energy shaping of mechanical systems via PID control and extension to constant speed tracking},
	author={Romero, Jose Guadalupe and Ortega, Romeo and Donaire, Alejandro},
	journal={IEEE Transactions on Automatic Control},
	volume={61},
	number={11},
	pages={3551--3556},
	year={2016},
	publisher={IEEE}
}

@article{sandoval2008interconnection,
	title={Interconnection and damping assignment passivity—based control of the pendubot},
	author={Sandoval, Jes{\'u}s and Ortega, Romeo and Kelly, Rafael},
	journal={IFAC Proceedings Volumes},
	volume={41},
	number={2},
	pages={7700--7704},
	year={2008},
	publisher={Elsevier}
}

@article{donaire2016simultaneous,
	title={Simultaneous interconnection and damping assignment passivity-based control of mechanical systems using dissipative forces},
	author={Donaire, Alejandro and Ortega, Romeo and Romero, Jose Guadalupe},
	journal={Systems \& Control Letters},
	volume={94},
	pages={118--126},
	year={2016},
	publisher={Elsevier}
}

@article{harandi2023stabilization,
	title={Stabilization of robots with actuator constraints via interconnection and damping assignment},
	author={Harandi, M Reza J and Namvar, Mehrzad and Taghirad, Hamid D},
	journal={IEEE Transactions on Control Systems Technology},
	volume={31},
	number={6},
	pages={2945--2952},
	year={2023},
	publisher={IEEE}
}

@article{j2021bounded,
	title={Bounded inputs total energy shaping for a class of underactuated mechanical systems},
	author={J. Harandi, M Reza and Taghirad, Hamid D and Molaei, Amir and Guadalupe Romero, Jose},
	journal={International Journal of Robust and Nonlinear Control},
	volume={31},
	number={18},
	pages={9267--9281},
	year={2021},
	publisher={Wiley Online Library}
}

@inproceedings{santibanez2005control,
	title={Control of the inertia wheel pendulum by bounded torques},
	author={Santibanez, Victor and Kelly, Rafael and Sandoval, Jesus},
	booktitle={Proceedings of the 44th IEEE Conference on Decision and Control},
	pages={8266--8270},
	year={2005},
	organization={IEEE}
}

@article{ortega2002stabilization,
	title={Stabilization of a class of underactuated mechanical systems via interconnection and damping assignment},
	author={Ortega, Romeo and Spong, Mark W and G{\'o}mez-Estern, Fabio and Blankenstein, Guido},
	journal={IEEE transactions on automatic control},
	volume={47},
	number={8},
	pages={1218--1233},
	year={2002},
	publisher={IEEE}
}

%
%

\end{document}